\documentclass[12pt]{article}

\usepackage[english]{babel}

\usepackage[a4paper,top=3cm,bottom=3cm,left=3cm,right=3cm,marginparwidth=1.75cm]{geometry}

\usepackage{amsmath}
\usepackage{graphicx}
\usepackage[colorlinks=true, allcolors=blue]{hyperref}

\title{Assessment of Analog Time Multiplexing in SDM Digital to Analog Converters}
\author{Alfredo P. Vega-Leal and José L. Mora}

\begin{document}
\maketitle

%
\begin{abstract}
Analog multiplexing for sigma delta modulated Digital to Analog Converters has been recently proposed as a means of achieving robustness. This preprint analyses said scheme via simulations. The main limitation introduced by the proposed architecture comes from mismatch in the DACs’ gain, which can drastically impact performances. A new technique of dynamic elements matching is proposed here to overcome this problem. 
\end{abstract}

%
\section{Introduction}
Sigma Delta Modulation (SDM) is receiving increasing attention in the implementation of modern communication standards, both in the design of transmitters and receivers. Wideband  transmitters require the digital-to-analog converter (DAC) to be rated at a very high frequency \cite{tanio2017fpga,colodro2010spectral}. 

In SDM-based Digital to Analog Converters (DAC)  speed constraints of the technology impose some limitations. If moderate oversampling ratio (OSR) values are chosen, moderate dynamic range (DR) values can be achieved; on the other hand, if a high OSR were used, the signal bandwidth would be low \cite{he2025100,candy1991oversampling}. Time-Interleaving (TI) has been proposed to alleviate speed constraints by trading off complexity and speed  \cite{khoini1997time,colodro1996cellular,kozak2000efficient}. The TI modulator can be derived from the classical one by replicating a basic modulator in $M$ parallel paths clocked at the rate $f_L=1/T_L$, such that the effective sampling rate is $f_H = M f_L =1/TH$. Therefore, the effective OSR is $M$ times greater than that of the basic modulator. Thanks to the current scale of integration, the higher complexity introduced by a TI implementation does not represent a significant problem in terms of cost, power consumption, and silicon area

The Error-Feedback (EF) structure is commonly used in the digital implementation of SDM based DACs \cite{candy1991oversampling,arahal2021adaptive}. In this architecture, clocked at the high rate $f_H$, the quantization error $E(z)$ is fed back to the input through the filter $H(z)$. Thus, the input is directly transferred to the output while the quantization error is affected by the Noise Transfer Function (NTF),

\begin{equation}
    Y(z) = X(z) + NTF(z) E(z)
\end{equation}

The loop filter is designed so that the NTF is a high pass filter

\begin{equation}
    NTF(z) = 1 + H(z) = (1 - z^{-1})^L/D(z)
\end{equation}

where $L$ is the order of the SDM and $D(z)$ is equal to one for $L = 1$ or $L=2$. If $L>2$, $D(z)$ must be a polynomial of order $L$ that stabilizes the loop \cite{colodro2014linearity,candy1991oversampling}.

Using the filtering approach \cite{kozak2000efficient,arahal2016harmonic,pham2008time,colodro2010continuous}, the architecture in Fig. 1 can be converted to its TI equivalent architecture (Fig. 2). The delayed replicas of the input $X(z)$ are separated in $M$ different paths and downsampled to the low rate $f_L$. The filtering is performed at the low rate by means of the block filter $\Bar{H}(z)$. In order to rebuild the high-rate output $Y(z)$, the outputs of the block filters are aligned in time by the up-samplers and the unit delays, and time multiplexed by the adders. The block filter is calculated from the polyphase decomposition terms of $H(z)$, and the output $Y(z)$  is a $M$-unit delayed replica of the previous case.

The added complexity introduced by the TI SDM based DAC is acceptable when the speed requirements are compromised by technological limitations \cite{colodro2003multirate}. When the technology prevents the design of the digital SDM at the rate $f_H$, TI is a solution. Unfortunately, the DAC used for converting the digital output $Y(z)$ to its analog counterpart$ V(w)$ becomes the limiting block in Fig. 2, since it still works at the high rate $f_H$. In order to overcome this bottleneck, the time multiplexing of the block filter outputs can be performed in the analog domain, so that the high speed DAC can be replaced by $M$ DACs in parallel clocked at the low rate $f_L$. These DACs are activated by $M$ clocks out of phase with each other the amount $2 \pi/M$. These phase shifts implement the unit delays shown in the time multiplexer. Finally, the DACs’ outputs are summed by means of an analog adder.

Parallel DACs working in TI mode have traditionally been used in Nyquist converters \cite{cheng2011nyquist,colodro2020open}. The increase of the effective conversion rate removes the closest Nyquist images and facilitates the design of the reconstruction filter. Unfortunately, errors in the DACs’ gains and among the parallel analog paths can deteriorate the performances of the system. These errors are produced by mismatch among electronic components. They can be mitigated by a careful layout \cite{deveugele2004parallel,colodro2009analog}, randomization in the use of DACs or digital correction \cite{hovakimyan2019digital}.

TI DACs can also be used in SDM based DACs to relax the speed requirements  \cite{arahal2024multi,mccue2016time}. Unfortunately, unlike the Nyquist converter, the signal at the input of the TI SDM based DAC has a large amount of quantization noise at high frequencies. Non-ideal effects can demodulate this high-frequency noise into the signal band. On the other hand, the inherent filtering of analog multiplexing can achieve greater robustness against timing skew \cite{colodro2022time}.

%
\section{Analyzed DAC Concept}

The output stage of the proposed SDM based modulator is depicted in Fig. 3.a. The outputs of the TI SDM are converted to the analog domain through a battery of single-bit and non-return to zero (NRZ) DACs. Unlike the DACs in Fig. 1 and Fig. 2, which are clocked at the high rate $f_H$ (with pulse width $T_H$), the DACs in Fig. 3.a generate their outputs at the low rate $f_L$ using NRZ pulses of width $T_L = 4 T_H$. The conversion cycle starts at the rising edge of the TL -period clocks cp(t) with $p = 0, 1, ..., M-1$. The unit delays shown in Fig. 2 are performed by means of shifting the relative phase of the parallel DACs. If the phase reference is the sampling time of any of the low rate signals $Y_p(z)$, the rising edge of $c_p(t)$ is delayed $p TH$ seconds.

\begin{figure}[ht]
    \centering
    \includegraphics[width=12cm]{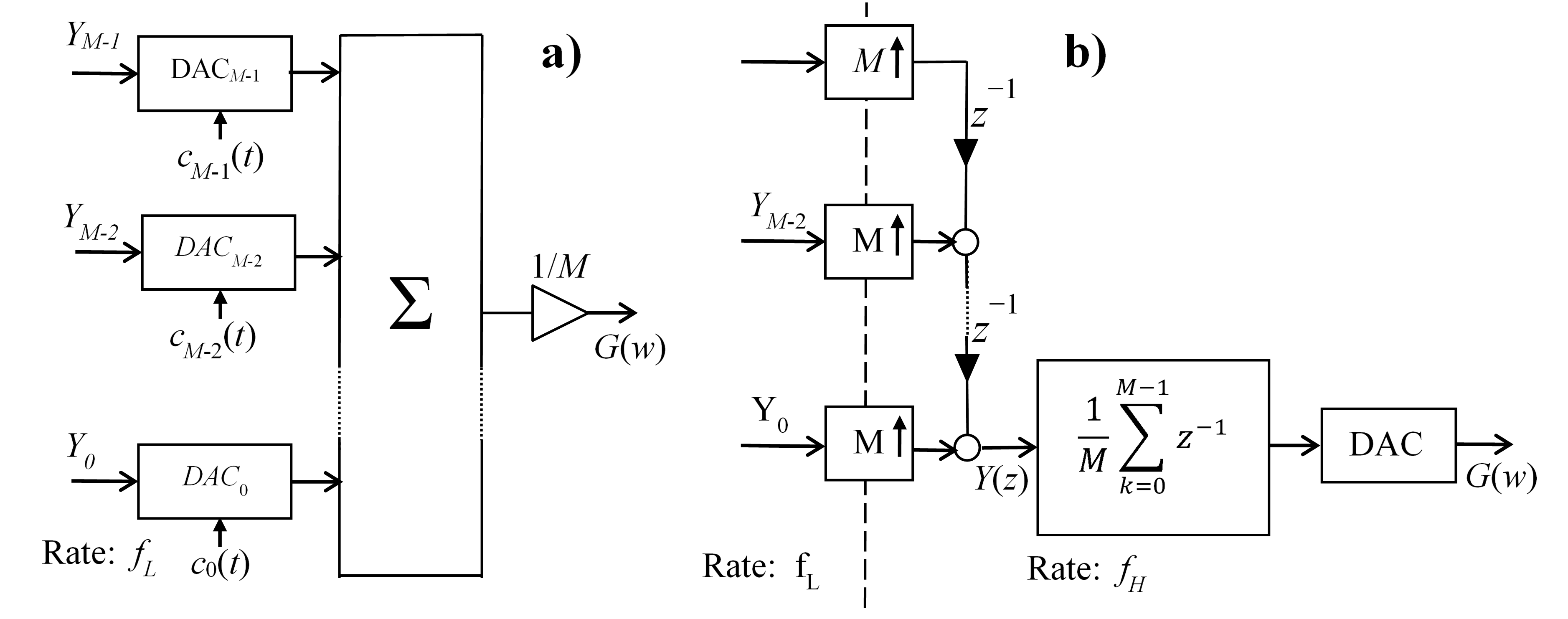}
    \caption{a) Output stage of the SDM based DAC and b) the equivalent discrete-time model.}
\end{figure}

%
\section{Effects of mismatches}
The first set of simulations is devoted to analyzing the effect of mismatches in discrete elements due to fabrication tolerances.

\subsection{Dynamic element matching}
Dynamic element matching (DEM) for the SDM based DAC is defined here. 
Every single DAC in Fig. 3a converts the logic levels 0 and 1 to the voltage levels $_VL$ and $V_H$, respectively. Mathematically, the $m$-th DAC, $m=0, 1, ..., M-1$, can be modelled as a linear stage with gain and DC offset: 
\begin{equation}
  V_{DAC,m} = g_m y_m + V_{off,m}    
\end{equation}

Due to tolerances in components in electronic implementations, the output voltages are also deviated from their ideal values and the gains ${ g_m }$ and DC offsets ${ V_{off,m} }$ take different values from one DAC to another. This mismatch between elements can also occur in the gains of the paths from $m$-th input of the adder in Fig. 3.a to its output. The value of this last gain can be included in the $m$-th DAC gain, without loss of generality.

Element mismatch is a very well-known issue in the design of multibit SDMs, where the SDM output has to be converted to analog through a multibit DAC. The most common topology uses a unit-element DAC. A $d$-bit binary weighted digital value, $x$, is converted by means of thermometric coding into a $D$-bit code, with $D = 2^d-1$, in such a way that
\begin{equation}
  x = \sum_{k=0}^{D-1} d_k
\end{equation}
\noindent with $d_k=1$ if $k<x$ and 0 otherwise. Now, the thermometric coded signal is converted to analog using $D$ unitary elements or single-bit DACs. For instance, if the discrete-time model had to be implemented, the DAC in Fig. 3b would be multibit, with an effective number of bits of $d=log2(M+1)=2.32$ for $M =4$. Unfortunately, the accuracy of the single-bit DAC gains must be as high as the resolution achieved by the SDM.

Data Weighted Averaging (DWA) is one of several methods to cope with the above problem. The basic operating principle of a SDM is the noise shaping; that is, the error (noise) produced by a low-resolution quantizer are filtered by a high-pass transfer function in such a way that this error is moved from the signal band to the high frequencies. In a similar way, by means of DWA, it is possible to shape the errors produced by the mismatch between DACs using a first-order high-pass transfer function. The working principle of DWA is based on element rotation: a) in each conversion period, the number of consecutive elements activated will be the number of high logic levels (ones) in the thermometric code; and b), the next elements to be activated succeed the previously ones according to the established order of rotation.

Unfortunately, the application of this technique to the proposed architecture is not straightforward. For instance, if the DAC shown in the DT model (Fig. 3b) has to be implemented using a unit-element structure, it would be necessary to have $M$ single-bit DACs rated at the high frequency $f_H$  operating in a synchronous way. On the other hand, the implementation of the architecture in Fig. 3a requires $M$ DACs rated at the low-frequency but, as explained above, operating with a relative phase shifting of $2 \pi/M$ each of the other. 
In the proposed DWA variant, the activation sequence of DACs in Fig. 3a is illustrated with an example in Fig. 8. For simplicity, instead of showing the four low-rate signals ${y_r, r =0, 1, 2 and 3}$, the multiplexed high-rate signal $y(n)$ is chosen (4). The DACs labelled “duty” take the high value $V_S$. On the other hand, those labelled “idle” take the low value $-V_S$. The pointer points to the DAC that will have to be activated when the next high logic level arrives. When the pointer is updated from the state 3, it will return to 0.

\begin{figure}[ht]
    \centering
    \includegraphics[width=9cm]{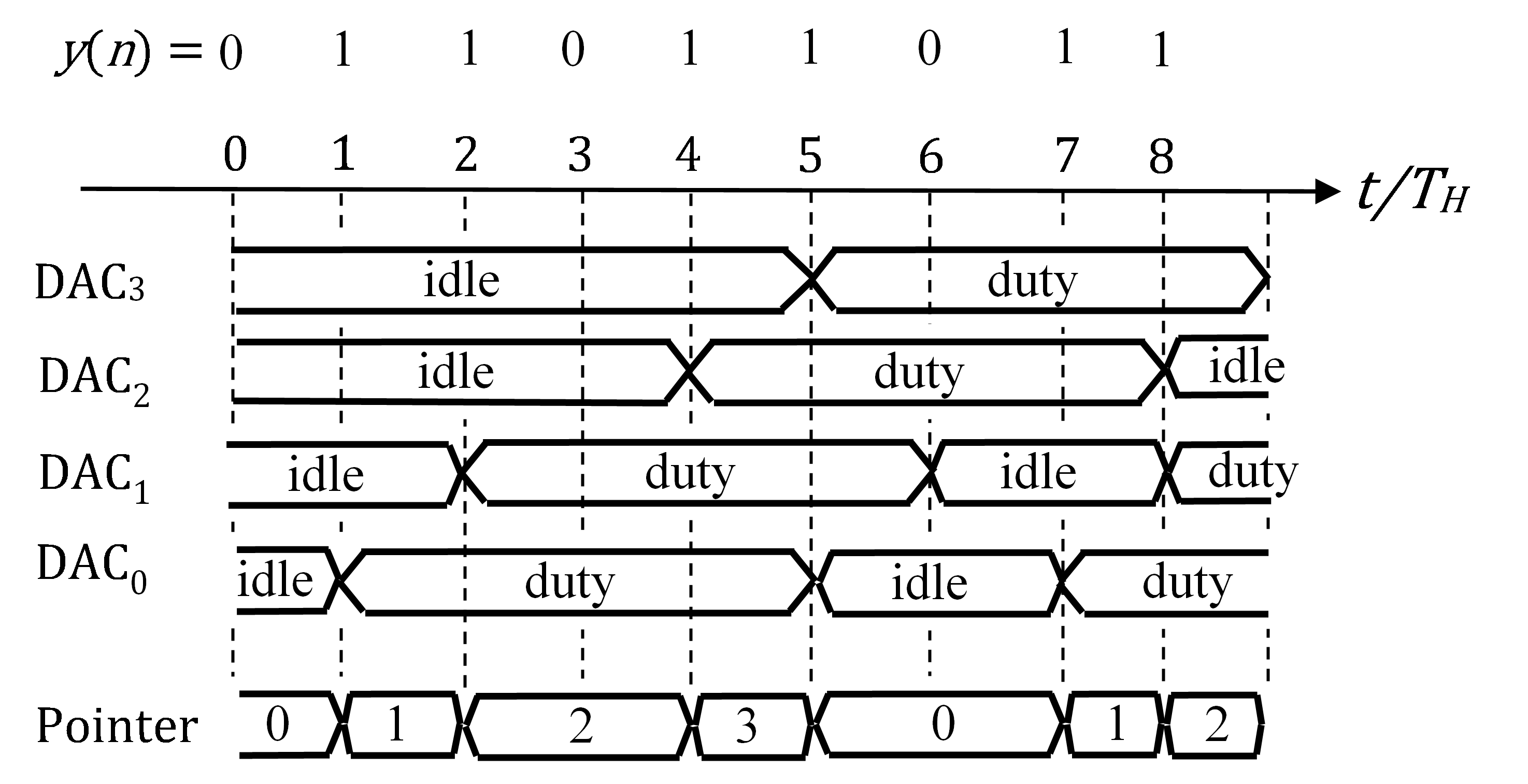} 
    \caption{Example of the activation sequence of the DACs.}
\end{figure} 

A possible implementation of this DWA variant is depicted in Fig. 9 (only for one of the $M$ DACs). As shown in Fig. 8, the pointer (not shown in Fig. 9) can be implemented using an up counter, synchronously clocked with a clock of frequency $f_H$, enabled every time that $y(n)=1$. The $m$-th DAC is clocked at the high rate $f_H$, but the pulse of width $T_L=M T_H$ is generated forcing its digital input at a high level during M cycles. This can be achieved by means of a down counter from 4 to 0. When the pointer points to the $m$-th DAC and $y(n)=1$, the counter is set to 4. As long as the counter value is greater than 0, the output of the 3-input AND gate is high, the $r$-th DAC is activated, and the counter is enabled for down counting.

\begin{figure}[ht]
    \centering
    \includegraphics[width=9cm]{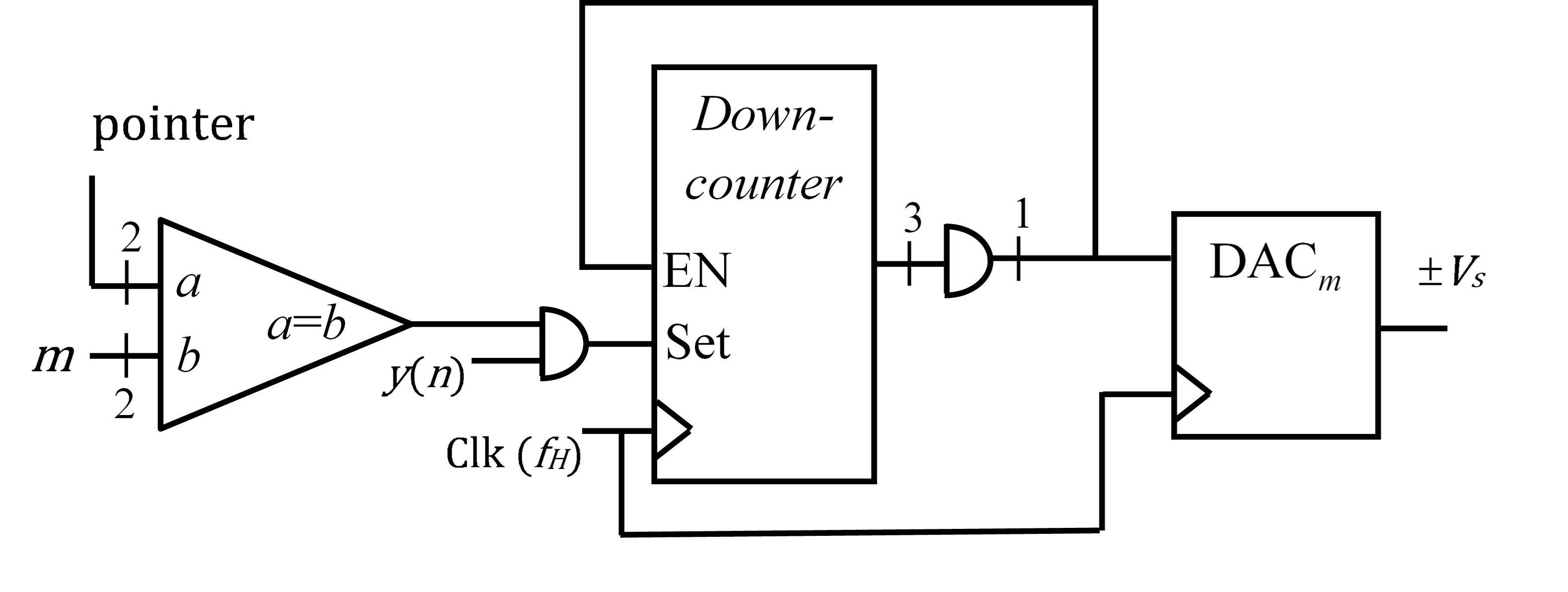} 
    \caption{Implementation of the DWA for the $m$-th DAC.}
\end{figure} 

Modern CMOS technologies easily achieve a matching between elements of around 1 \%, although a value as low as 0.1 \% can be achieved when the circuit layout is carefully designed and with the help of compensation techniques. Simulations reveal that, for a SNDR maximum of around 70 dB, the proposed architecture with DWA can tolerate up to a 5.8 \% of mismatch with an averaging decrease in the SNDR no greater than 3 dB. For the sake of illustration, for an input amplitude $A=0.707$ (-3 dBfs), the spectra of $G(w)$  are depicted in Fig. 10. The gain and DC offset deviations has been chosen from a uniformly random distribution, ranging from -0.09 to +0.09 (with standard deviation of 5.2 \%). 

The selected values are: $\{ g_m \} = \{ 1.07, 0.93, 0.98, 0.96 \}$ and $\{V_{off,m}\}= \{ 0.05,$ $ -0.01, 0.07, -0.06 \}$. The SNDR is 69.7 dB for the ideal case and 37.8 dB with the selected mismatch. When DWA is applied, the SNDR is recovered to the value of 67.0 dB. Let us note from Fig. 10 that the effect of mismatch is to demodulate quantization noise from the high frequency region to the signal band. When the DWA is applied, this demodulated noise can be appreciated too, but it does not degrade the performances. The bottom spectrum of the Fig. 10 falls 20 dB/dec. As expected, the mismatch errors are shaped by a first-order transfer function.

\begin{figure}[ht]
    \centering
    \includegraphics[width=15cm]{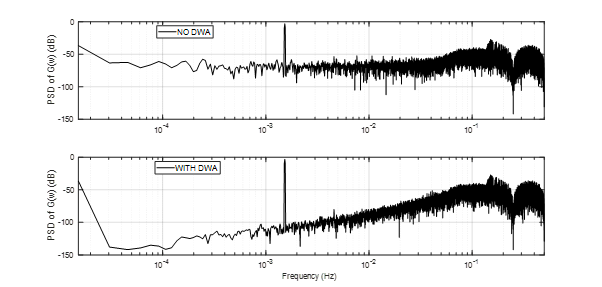} 
    \caption{Spectra of the proposed architecture with a mismatch of 5.2\%.}
\end{figure} 

\subsection{The case of multibit DACs}
The aforementioned algorithm can be straightforwardly extended to the case of a SDM based DAC with $d$-bit quantizers. Now, the $M$ DACs in Fig. 3a must also be of $d$ bits and each of them could be implemented using $D =2^d -1$ unit elements. Therefore, the total number of unit elements (or single-bit DACs) will be $MD$. Every set of $D$ unit elements could be assigned to one DAC and DWA could be independently applied to each of the $M$ DACs. But even so, the algorithm would not work because the total gain and DC offset of every independent DAC would be, under mismatch, different from each other. Therefore, the $M (2^d-1)$ unit elements must be shared among all DACs. 

Let $y(n)$ be coded in natural binary; $y \in {0, 1, ..., (2^d-1)}$. Then the pointer has to sweep from 1 to $M (2^d-1)$ and must be increased every clock cycle by the amount $y(n)$. Again, the algorithm works as long as the elements are chosen in a rotatory and succeeding that guarantees that every element is activated the same number of times on average.  The logic can again be implemented as the circuit in Fig. 9 with minor differences. 

The DR graphs are now shown in Fig. 11 for an OSR=32 with quantizers of 3 bits, giving a total number of unit elements equal to $4 \times 7=28$ ($M=4$). The mismatch is of 4 \%: gain values are unity plus a random number uniformly distributed from -0.07 to +0.07. As it can be seen, even for this high mismatch, the algorithm still works.

\begin{figure}[ht]
    \centering
    \includegraphics[width=12cm]{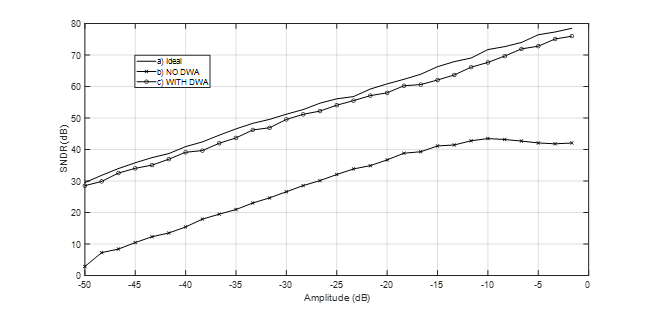} 
    \caption{DR graph for the 3-bit quantizer, $M=4$ and a 4 \% mismatch: a) ideal, b) with mismatch and without DWA, and c) with mismatch and DWA.}
\end{figure} 

%

%
\section{Errors produced by the shape of the pulses}
Another source of errors is the shape of the DAC pulses. This is especially important at high frequencies, where the rising and falling times are a non-negligible fractions of the bit period. SDM based DACs with NRZ pulses are more sensitive to deviations from the ideal rectangular shape than those with RZ pulses. As shown in Fig. 12, the waveforms can vary depending on the sequence of bits for the case of NRZ pulses.  For instance, the three high-logic levels in Fig. 12 are converted with different waveforms and, consequently, with different values of integrated area. The actual waveform depends on the signal level (sequence of bits) producing a non-linear effect. 

As a consequence, the SNDR is degraded because the high frequency quantization noise is folded to the signal band and harmonic distortion can be produced. On the other hand, the waveforms can be maintained the same using RZ pulses. Nevertheless, these last pulses require more severe constraints in terms of speed. 

\begin{figure}[ht]
    \centering
    \includegraphics[width=9cm]{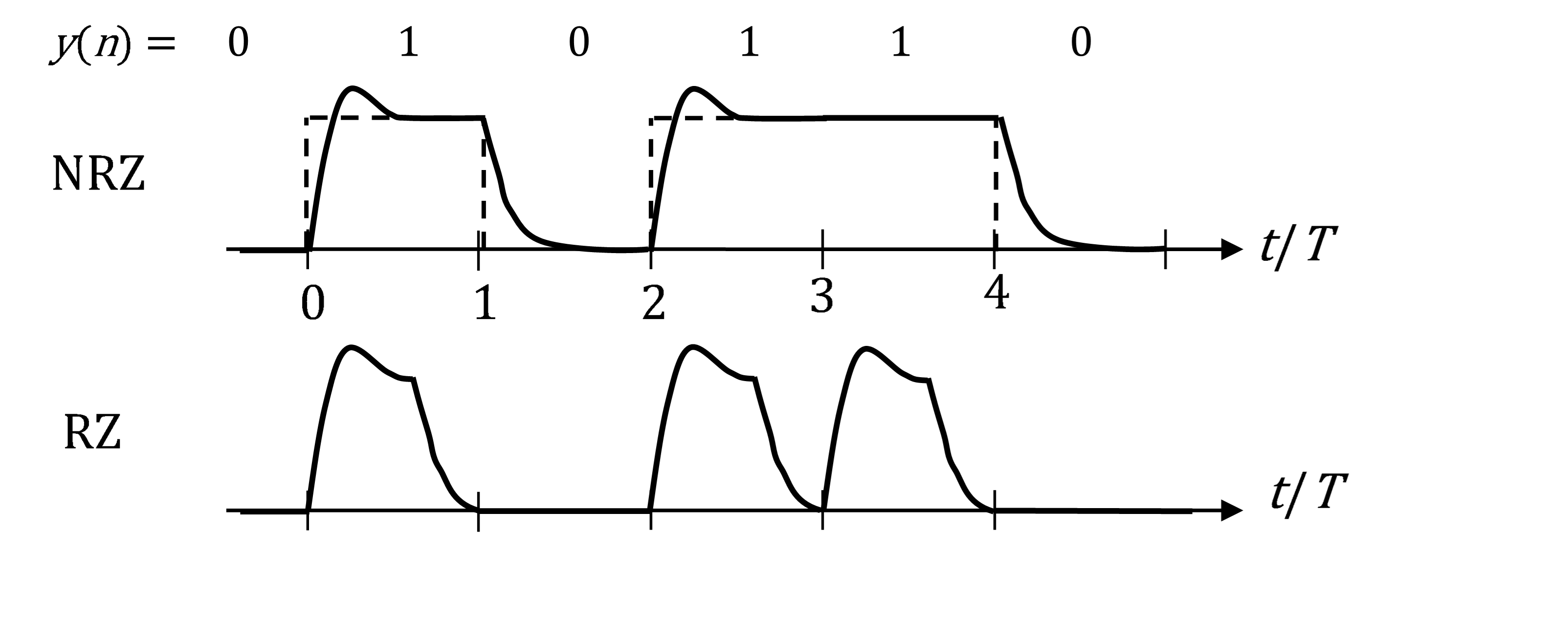} 
    \caption{Waveforms at the DAC output for NRZ and RZ pulses. Ideal pulses with dashed line.}
\end{figure} 

\begin{figure}[ht]
    \centering
    \includegraphics[width=9cm]{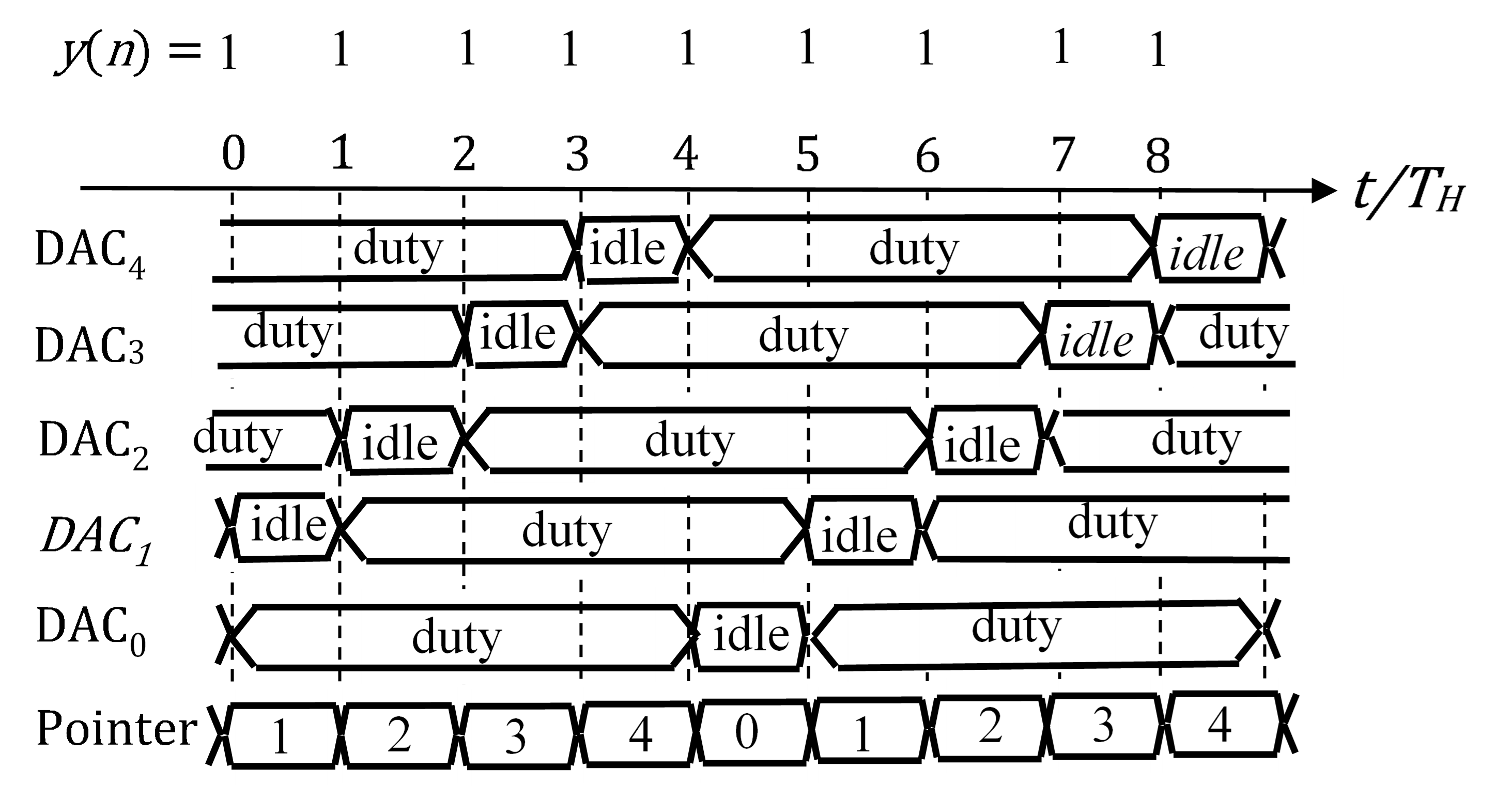} 
    \caption{Activation pattern with 5 DACs for M=4 and a long ones’ sequence.}
\end{figure} 

Due to the pulses of the DACs in Fig. 3a are $M$ times wider  for the same rising and falling times, the errors in the integrated area will be  $M$ times lower, and consequently, it is expected that the architecture in Fig. 3a will be $20 \log_{10}(M)$ dB more robust than that of Fig.1. An additional improvement can be achieved if the pulses are forced to return to zero. This can be carried out if M+1 DACs are used. For $M=4$, the activation sequence is as shown in Fig. 8, but adding one additional DAC. Since the number of signals to be converted is 4 and the number of DACs is 5, there will always be, at any instant, one or more DACs in idle state and every DAC pulses will be forced to return to zero after 4 clock cycles.  For instance, as shown in Fig.13, a long sequence of ones produces the highest activity, the activation sequence is periodic, and each 
DAC alternates 4 duty cycles with 1 idle cycle.
The architectures of Fig. 1 and Fig. 3a have been simulated. The analog multiplexing of Fig. 3a has been simulated with 4 and 5 DACs.  In the last case, as explained in the previous paragraph, with pulses that return to zero and width $T_L=4 T_H$. The rectangular pulses have been shaped using the circuit in Fig. 14. Two effects are modeled: first, the slew rate (SR) or limitation in the maximum rate of change; and second, a first-order exponential change with a time constant $\tau$. As long as $|v_e| < V_o$ , the saturation at the integrator input does not work and the circuit behaves as a linear first-order system. Nonetheless, if $Vo < 2V_S$, at the instants where the input changes, $|ve| =2V_S$, the saturation works and the output goes into SR: $|v_o(t)| = SR t$ , where $SR=Vo /\tau $ . SR is a nonlinear effect that can produce harmonic distortion and demodulation of the high-frequency quantization noise of the SDM based DAC.

\begin{figure}[ht]
    \centering
    \includegraphics[width=9cm]{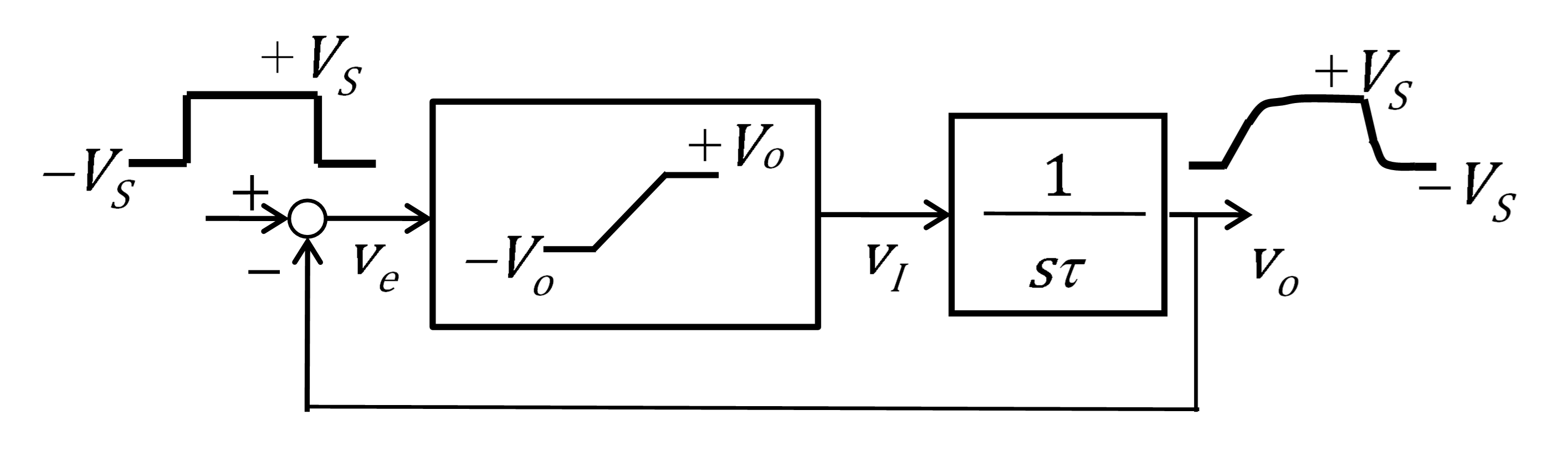} 
    \caption{Modeling of the shape of the DAC pulse.}
\end{figure} 

The PSDs in the signal band for an input amplitude of -3 dB are shown in Fig. 15. The chosen nominal values of the SR and the time constant are $SR =1.5 V_s f_H$ , and  $\tau =0.5/f_H$, respectively. In addition, a 5 \% deviation from the nominal values has been taken to differentiate the parameters of the down-to-up ( $\tau_p$ and $SR_p$) and up-to-down ( $\tau_n$ and $SR_n$)transitions for the DAC in Fig. 1. 

The parameters of the DACs in Fig. 3a have been obtained from a uniform random distribution of standard deviation of 5 \%. The first four parameter sets for the case of 5 DACs are the same as the four parameter sets for the case of 4 DACs. 

\begin{figure}[ht]
    \centering
    \includegraphics[width=15cm]{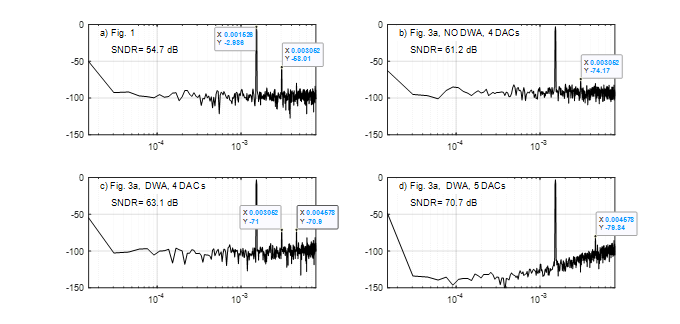} 
    \caption{PSD (dB) versus normalized frequency (Hz), $f_H=1 $Hz, in the signal band. The SNDR in the ideal case (rectangular pulses) is 70.7 dB.}
\end{figure} 

Some conclusions can be drawn. Let us note that there is no performance loss in the case that 5 DACs are used and the DWA algorithm is applied (graph d). The achieved SNDR, 70.7 dB, is the same as in the ideal case (rectangular waveforms in the DACs). Moreover, the RZ effect, unlike the rest of cases, avoids the generation of harmonic distortion, since there is no second harmonic, and the third one has the same amplitude, -79.8 dB, as the third harmonic in the ideal case (that is, this harmonic is generated by the digital SDM and not by the DACs). The expected improvement of 12 dB due to the use of pulses 4 times wider can also be observed from the differences between the SNDRs of graph a) and graphs b) - c). The same can be said for the amplitude of the second harmonic.

%
\section{Conclusion}
The main drawback of the analyzed architecture is the element mismatch among the parallel paths going from the input of each DAC to the output of the system. Fortunately, a variant of the DWA algorithm proposed here can overcome this problem. As shown, the SDM based DAC is insensitive to a mismatch as high as 5.7 \%, a value much larger than the practical value of 1 \% that can easily be achieved with today’s technologies. The implementation of the DWA variant is not resource-consuming, since it only requires to replicate M  times the simple circuit shown in Fig. 9, which comprises a 2-bit comparator, a 3-bit counter and two AND gates for $M=4$. Despite the fact that the DWA variant has to be rated at the frequency $f_H$, this is not a limitation for two reasons. Firstly, due to the simplicity of the circuits in Fig. 9, there is not a large combinational logic whose long delay can limit the operating frequency. Secondly, if this were the case, since the circuit in Fig. 9 is not in a closed loop where an added delay could compromise stability, it is possible to break the combinational paths by including flip-flops. For instance, between the comparator output and the succeeded AND gate or between the internal combinational logic of the comparator. These flip-flops would only influence the latency of the SDM based DAC, which is not a critical issue. Finally, it has been shown that the DWA variant can also be applied to the case when the DACs is multibit.

%
\bibliographystyle{unsrt}
\bibliography{sample}

\end{document}